\documentclass[12pt]{article}
\usepackage{epsf, cite, amssymb}
\usepackage{epsfig}

\makeatletter
\renewcommand\section{\@startsection {section}{1}{\z@}%
                                   {-3.5ex \@plus -1ex \@minus -.2ex}
                                   {2.3ex \@plus.2ex}%
                                   {\normalfont\large\bfseries}}
\renewcommand\subsection{\@startsection{subsection}{2}{\z@}%
                                     {-3.25ex\@plus -1ex \@minus -.2ex}%
                                     {1.5ex \@plus .2ex}%
                                     {\normalfont\bfseries}}
\makeatother

\let\non\nonumber

\let\a=\alpha\let\b=\beta

\newcommand{\bea}{\begin{eqnarray}}
\newcommand{\eea}{\end{eqnarray}}
\newcommand{\be}{\begin{equation}}
\newcommand{\ee}{\end{equation}}

\newcommand{\Z}{{\mathbb Z}}

\newcommand{\e}{\epsilon}

\newcommand{\rr}{\rightarrow}
\newcommand{\m}{\mu}

\newcommand{\p}{\partial}

\newcommand{\tr}{{\rm Tr}}

\newcommand{\C}[1]{$(\ref{#1})$}

\def\IZ{\relax\ifmmode\mathchoice
{\hbox{\cmss Z\kern-.4em Z}}{\hbox{\cmss Z\kern-.4em Z}}
{\lower.9pt\hbox{\cmsss Z\kern-.4em Z}} {\lower1.2pt\hbox{\cmsss
Z\kern-.4em Z}}\else{\cmss Z\kern-.4em Z}\fi}
\def\IR{\relax{\rm I\kern-.18em R}}

\def\one{{\hbox{ 1\kern-.8mm l}}}

\def\tr{{\rm tr\,}}

\newlength{\bredde}
\def\slash#1{\settowidth{\bredde}{$#1$}\ifmmode\,\raisebox{.15ex}{/}
\hspace*{-\bredde} #1\else$\,\raisebox{.15ex}{/}\hspace*{-\bredde}
#1$\fi}

\newsavebox{\zzzbar}
\sbox{\zzzbar}
  {\setlength{\unitlength}{0.9em}
  \begin{picture}(0.6,0.7)
  \thinlines
  \put(0,0){\line(1,0){0.6}}
  \put(0,0.75){\line(1,0){0.575}}
  \multiput(0,0)(0.0125,0.025){30}{\rule{0.3pt}{0.3pt}}
  \multiput(0.2,0)(0.0125,0.025){30}{\rule{0.3pt}{0.3pt}}
  \put(0,0.75){\line(0,-1){0.15}}
  \put(0.015,0.75){\line(0,-1){0.1}}
  \put(0.03,0.75){\line(0,-1){0.075}}
  \put(0.045,0.75){\line(0,-1){0.05}}
  \put(0.05,0.75){\line(0,-1){0.025}}
  \put(0.6,0){\line(0,1){0.15}}
  \put(0.585,0){\line(0,1){0.1}}
  \put(0.57,0){\line(0,1){0.075}}
  \put(0.555,0){\line(0,1){0.05}}
  \put(0.55,0){\line(0,1){0.025}}
  \end{picture}}

\newcommand{\ena}{\end{eqnarray}}
\newcommand{\beqa}{\begin{eqnarray}}
\newcommand{\eeqa}{\end{eqnarray}}

\newcommand{\eq}[1]{(\ref{#1})}

\renewcommand{\b}{\beta}

\def\a{\alpha}
\def\b{\beta}

\def\e{\epsilon}

\def\m{\mu}

\newcommand{\Dslash}{\ensuremath \raisebox{0.025cm}{\slash}\hspace{-0.32cm} D}

\begin{document}
\begin{titlepage}

\begin{center}

\today \hfill     EFI-05-21, UCI-TR-2005-47, ITFA-2006-02

\vskip 2 cm
{\Large \bf Effective Dynamics of the Matrix Big Bang}\\
\vskip 1.25 cm { Ben Craps$^{\,a, \, b}$\footnote{email address:
 Ben.Craps@vub.ac.be}, Arvind Rajaraman$^{\,c}$\footnote{email address: arajaram@uci.edu} and
 Savdeep Sethi$^{\,d}$\footnote{email address:
 sethi@theory.uchicago.edu}
}\\
{\vskip 0.5cm $^{a}$ Theoretische Natuurkunde, Vrije Universiteit Brussel and\\
The International Solvay Institutes\\ Pleinlaan 2, B-1050
Brussels, Belgium\footnote{present address}\\
\vskip 0.5cm $^{b}$ Instituut voor Theoretische Fysica,
Universiteit van Amsterdam, Valckenierstraat 65, 1018 XE
Amsterdam,
The Netherlands  
\\
\vskip 0.5cm $^{c}$ 
Department of Physics and Astronomy, University of California, Irvine, CA 92697, USA
\vskip 0.5cm $^{d}$ Enrico Fermi Institute,
University of Chicago, Chicago, IL 60637, USA\\}

\end{center}

\vskip 2 cm

\begin{abstract}
\baselineskip=18pt

We study the leading quantum effects in the recently introduced Matrix Big Bang model. This amounts to a study of supersymmetric Yang-Mills theory compactified on the Milne orbifold. We find a one-loop potential that is attractive near the Big Bang. Surprisingly, the potential decays very rapidly at late times, where it appears to be 
generated by D-brane effects. 
Usually, general covariance constrains the form of any effective action generated by renormalization group flow. However, the form of our one-loop potential seems to violate these constraints in a manner that suggests a connection between the cosmological singularity and long wavelength, late time physics.  

\end{abstract}

\end{titlepage}

\pagestyle{plain}
\baselineskip=19pt
\section{Introduction}

One of the basic lessons emerging from recent work on cosmological singularities is the need to understand interacting field theories which are not Lorentz invariant. 
These field theories describe physics in cosmological settings via holography. The breaking of Lorentz invariance is inevitable since cosmological space-times distinguish time. Much of the past work on field theory in curved space-times focuses on free fields (see, for example,~\cite{Birrell:1982ix}). The inclusion of interactions turns out to be remarkably non-trivial. This can be understood intuitively: as time evolves from the Big Bang, 
the wavelength of an excitation red-shifts. Therefore, the notion of a high or low-energy excitation becomes time-dependent. Correspondingly, the use of renormalization group flow to define ``low-energy'' dynamics becomes subtle.  

Our goal in this work is to study the dynamics of the particular Big Bang model introduced in~\cite{Craps:2005wd}\ based on Matrix theory~\cite{Banks:1996vh}. For a sampling of recent work on holography and cosmological singularities, see~\cite{holo, Li:2005sz, Li:2005ti, Kawai:2005jx, Hikida:2005ec, Das:2005vd, Chen:2005mg, She:2005mt, Chen:2005bk, Ishino:2005ru, Hikida:2005xa, Robbins:2005ua, Silverstein:2005qf, She:2005qq}. The space-time theory is type IIA string theory in flat space with a non-trivial dilaton, 
\be
\phi = - Q X^{+}. 
\ee
The Big Bang occurs as $X^{+}\rightarrow -\infty$ where the theory is strongly coupled. At late times, perturbative string theory becomes a good description. 

After compactifying $X^{-} \sim X^{-} + R$, we are led to a non-perturbative definition of string theory on this background given in terms of maximally supersymmetric Yang-Mills in two dimensions compactified on the Milne orbifold. This should be contrasted with studies where the Milne orbifold appears as a string theory target space~\cite{Cornalba:2002fi, Nekrasov:2002kf, Craps:2002ii, Berkooz:2002je, Pioline:2003bs, Craps:2003ai, Berkooz:2004re, Berkooz:2004yy}.
At early times near the Big Bang, the model is perturbative Yang-Mills while at late times the model reduces to perturbative strings on the light-like linear dilaton background.  

In section~\ref{matrixfermions}, we will directly check that the Matrix model fermions are indeed fermions on Milne space. In terms of the orbifold description of Milne space, these fermions have unconventional periodicity properties which are described in section~\ref{wavefunctions}. The upshot of this discussion is that the breaking of Poincar\'e invariance on the world-sheet is spontaneous in this model, in the sense that the field theory action is invariant under diffeomorphisms if the background metric is transformed as well. We should note that in other related Matrix models, the breaking of Poincar\'e invariance is not spontaneous; see, for example~\cite{Robbins:2005ua}. 

The question that we wish to address is how to describe the effective dynamics in our Matrix model. This involves understanding how to define effective dynamics in a time-dependent background, and then determining whether a potential  is generated in this model. 

We will find that the presence of a Big Bang is detectable at late times as an effective potential generated by infrared physics. This is a little surprising since the Big Bang is an ultraviolet phenomena. What is more strange is the form of the potential. The structure of the late time effective action appears to violate arguments following from general covariance which we will describe in the following section. We suspect that this will be a quite general phenomenon suggestively connecting infra-red physics to a cosmological singularity.   

The potential we find is attractive near the Big Bang and decays very rapidly at late times. This strongly suggests that the flat directions required in Matrix theory for a space-time interpretation are restored at late times despite the broken supersymmetry. We seem to avoid the kind of problems encountered in~\cite{Banks:1999tr}. This late time regime is where we expect perturbative string theory to be a good description so this rapid decay is unexpected good news. Given the extremely rapid late time decay, it might even be the case that higher loop corrections to the potential are suppressed despite the strong Matrix model coupling. If so, we predict the existence of a potential in string theory between gravitons with separation $b$ in the light-like linear dilaton background with leading asymptotic behavior
\be
 \int \sqrt{g} \, V_{eff}(b)  \sim   \int \sqrt{{b\over g_{s}}}  \exp\left(-{Cb\over g_s} \right),
\ee 
with constant $C$. 
The form suggests that the potential is generated by D-brane effects in this time-dependent background. 

\vskip 0.5cm
\noindent {\bf Note added:} During the completion of this project, we received~\cite{Li:2005ai}\ which discusses similar issues in a slightly different model. 

\section{Spontaneous Breaking of Poincar\'e Invariance}
\subsection{Some comments on general covariance}  

Consider a local quantum field theory in a background spacetime with metric $g$, such that the classical action is invariant under general coordinate transformations if the background metric is transformed appropriately. If the spacetime has non-trivial topology, we may need to specify choices like the spin structure for fermions and perhaps the topology of the fields. What can we say about the effective quantum dynamics? If there are no gravitational anomalies, then general coordinate transformations (accompanied by transforming the background metric) and local Lorentz transformations must be symmetries preserved under renormalization group flow. If for instance we assume that the field theory can be consistently coupled to gravity by making the metric dynamical, this must be the case.  

What we conclude is that the low-energy theory can be described by an expansion of the schematic form, 
\be\label{generalexp}
S_{{eff}} = \int \sqrt{\tilde{g}} \left\{ \p_{\mu} \Phi \p^{\mu} \Phi + V^{(0)}(\Phi) + R(\tilde{g}) V^{(1)} (\Phi) + \ldots \right\},
\ee
where $\Phi$ are light fields and $R(\tilde{g})$ is the Ricci scalar for a possibly renormalized metric $\tilde{g}$. All terms in this action, including omitted terms, must respect diffeomorphism invariance. 

In particular $V^{(0)}(\Phi)$ is metric independent and so can be computed using a flat metric. For a theory with extended supersymmetry, this potential vanishes. On the other hand,  $V^{(1)}(\Phi)$ can be non-vanishing on space-times with non-vanishing curvature even in supersymmetric theories.  

In our particular case, we are interested in maximally supersymmetric Yang-Mills compactified on the Milne orbifold. The Milne orbifold is obtained from Minkowski space,
\be\label{Mink} ds^2=-2d\xi^+d\xi^-, \ee by the boost
identification 
\be\label{boost} \xi^\pm\sim\xi^\pm e^{\pm 2\pi
Q\ell_s}. \ee 
In terms of new coordinates $\tau,\sigma$ defined by
\be\label{relation}
\xi^\pm={1\over\sqrt2Q}e^{Q(\tau\pm\sigma)}, 
\ee the metric reads 
\be\label{Milne} ds^2=e^{2Q\tau}(-d\tau^2+d\sigma^2) \ee 
and the identification is 
\be \label{id} \sigma\sim\sigma+2\pi\ell_s. \ee 
We can
further introduce light-cone coordinates 
\be
x^\pm={1\over\sqrt2}(\tau\pm\sigma), 
\ee in terms of which the
metric reads 
\be\label{Milnebis}
ds^2=e^{\sqrt2Q(x^++x^-)}(-2dx^+dx^-). \ee

Without the periodic identification~\C{id}, this space-time is equivalent to flat space (much like taking the Rindler slice in four dimensions). All curvatures are therefore vanishing everywhere except possibly at the fixed point $\tau = -\infty$ which corresponds to the Big Bang. 

This point is rather critical. Just from the structure of the metric, we see that the presence of the cosmological singularity is detectable by long wavelength physics at late times: namely, by detecting the presence of a large circle varying in time and locally vanishing curvatures. This same circle identification induces both the cosmological singularity and also supersymmetry breaking. 

What we will find by computation is that a potential is generated in this model because of infrared effects. One way to understand this is by viewing the theory as an orbifold compactification in terms of $(\xi^{+}, \xi^{-})$. In this frame, the circle identification is up to the action of a boost. This orbifold identification leads to a twisted boundary condition that depends on the spin of the particle. 

In particular, this twist leads to a mismatch in boson/fermion oscillator frequencies on the circle. Without the usual supersymmetric cancellation, a potential is generated. However, as we will describe, the mismatch in frequencies is very small at high energies but significant at low-energies. The corresponding potential is not localized at time $\tau=-\infty$ as one might expect from the general form of~\C{generalexp}. Rather the potential leads to rapidly decaying but observable effects even at late times.

This amounts to a strange intertwining of long distance/late time physics (effects from the large circle) and the cosmological singularity. In orbifold compactifications of flat space, these kinds of correlations seem unavoidable since the curvature is vanishing everywhere but the fixed point set. Unlike static orbifolds, all modes detect these effects in models with a cosmological singularity. For example, all modes detect the breaking of supersymmetry.

There are a few possible explanations that might reconcile the form of the potential we find with the structure expected from covariance. It is possible that the effective action is ill-defined on cosmological backgrounds. In section~\ref{effective}\ we will argue that the Wilsonian effective action is ill-defined on any space-time where the blue-shift can be arbitrarily large. However, at least in cases like the Milne orbifold, the 1PI effective action appears to be sensible.  It is also possible that on spaces with a non-trivial fundamental group, like the Milne orbifold, there might exist more diffeomorphism invariants that could be used to construct the potential.  

\subsection{Fermions in the matrix model}
\label{matrixfermions}

Our first task is to show that the fermions appearing in the Matrix Big Bang model are actually fermions on Milne space. This is important for two reasons: first to demonstrate that Lorentz invariance is spontaneously rather than explicitly broken. Second, to determine the choice of spin structure on the world-volume circle. 

Fermion couplings at the level of the DBI action have been analyzed in~\cite{Marolf:2003vf, Marolf:2004jb, Martucci:2005rb}. We need to study the type IIB background related by U-duality to the type IIA string with a light-like linear dilaton and a compact direction $X^{-} \sim X^{-} + R$. Let us consider the case of a single D-string wrapped on $x^{1}$ in this type IIB background 
\bea
ds^2& =& r e^{\epsilon Q x^+} \left\{ -2 dx^+ dx^- + \sum_{i=1}^8 (dx^i)^2\right\},\label{metricD1}\\
\phi &=& \epsilon Q x^+ + \log r\label{dilneg},
\eea
with the identification
\be
x^1 \sim x^1 + {2\pi\ell_s \over r}.
\ee
The parameter $r$ is defined to be
\be
r \equiv {\epsilon R \over 2\pi\ell_s}
\ee
and we identify the world-volume coordinates $(\tau,\sigma)$  with space-time coordinates as follows
\bea
x^1 &= &{1\over r}\sigma,\nonumber\\
x^+ &=& {1\over r}{\tau \over \sqrt2}.\label{gaugexplus}
\eea

After fixing the $\kappa$-symmetry and setting the gauge field strength to zero, the fermion couplings can be expressed in the form
\be\label{fermions}
S_{F} = {1\over 2\pi \ell_{s}^{2}} \int d\tau d\sigma e^{-\phi} \sqrt{\det(-g)} \, \bar\psi \Dslash \psi. 
\ee
The metric $g$ appearing in \C{fermions}\ is the pull-back to the world-volume of the space-time metric \C{metricD1}, 
\be
ds^{2} = {1\over r} e^{\e Q \tau \over \sqrt{2} r} \left\{ -d\tau^{2} + d\sigma^{2}\right\}.
\ee
The same is true for the Dirac operator. 
It is notable that a coupling to the gradient of the dilaton could have been present but vanishes by symmetry considerations~\cite{Martucci:2005rb}.

The prefactor multiplying the fermion bilinear in~\C{fermions}\ can be read from~\cite{Craps:2005wd}
\be
e^{-\phi} \sqrt{\det(-g)} = {1\over r^{2}}. 
\ee
We also need to compute the pull-back of the spin connection. To determine the spin connection, we make a choice of frame bundle
\be
e^{i} = \sqrt{r} e^{\e Q x^{+}/2} dx^{i}. 
\ee
With this choice, the spin connection pulled back to the brane has non-vanishing components
\be\label{spin}
\omega^{1-} = -{\e Q \over 2r} d\sigma, \qquad \omega^{-+} = {\e Q \over 2\sqrt{2} r} d\tau.  
\ee
It is easiest to view the fermion appearing in~\C{fermions}\ as a Majorana-Weyl fermion in $9+1$-dimensions. Under reduction to the world-volume of the D-string, the fermion decomposes into representations
\be
{\bf 1}_{+} \otimes {\bf 8}_{c} +  {\bf 1}_{-} \otimes {\bf 8}_{s}
\ee 
under the $Spin(1,1) \times Spin(8)_{R}$ symmetry group in $1+1$-dimensions. However, one can readily check that the spin connection~\C{spin} drops out of~\C{fermions}. 

In terms of flat space gamma matrices, the fermion kinetic term is then given by
\be\label{fermcoup}
e^{-\phi} \sqrt{\det(-g)} \, \bar\psi \Dslash \psi = {1\over r^{3/2}} e^{-{\e Q\tau \over 2\sqrt{2} r}} \bar\psi  \left( \gamma^{0} \partial_{\tau} + \gamma^{1} \partial_{\sigma} \right)\psi. 
\ee
Note we can always field redefine the fermions to remove the time-dependence in~\C{fermcoup}\ without introducing any new couplings. So from this abelian expression, we cannot determine whether the fermions really see the Milne orbifold because they are completely decoupled from the boson sector of the theory. 

What we require is the non-abelian generalization of~\C{fermcoup}. Fortunately, this can be determined in the following way. The fermion kinetic term is obtained by starting with non-abelian $9+1$-dimensional Yang-Mills coupled to the background metric. This amounts to including a $9+1$-dimensional gauge-field in $\Dslash$ and taking a trace. This covariant derivative is then pulled back to the world-volume of the brane to give both the fermion kinetic terms and the Yukawa couplings. 

On performing this procedure, we find the following quadratic fermion couplings,
\be
{1\over r^{3/2}}  {\rm tr} \left\{ e^{-{\e Q\tau \over 2\sqrt{2} r}}  \bar\psi  \left( \gamma^{0} D_{\tau} + \gamma^{1} D_{\sigma} \right)\psi  + e^{{\e Q \tau\over 2\sqrt{2} r }}\bar\psi \gamma^{i} \left[ X^{i},  \psi \right] \right\},
\ee
where the $\gamma^{i}$ are standard gamma matrices for $Spin(8)$. The scalar $X^{i}$ is in the adjoint representation of the $U(N)$ gauge group. We are now free to rescale $\psi$ so the fermion kinetic terms are canonical
\be\label{mixferm}
 {\rm tr} \left\{ \bar\psi  \left( \gamma^{0} D_{\tau} + \gamma^{1} D_{\sigma} \right)\psi  + e^{{\e Q \tau\over \sqrt{2} r }}\bar\psi \gamma^{i} \left[ X^{i},  \psi \right] \right\}.
\ee
Note that all the $\e$ factors cancel as we might expect. By comparison with~\cite{Craps:2005wd}, it is easy to see that this is precisely the right coupling needed for fermions on the Milne orbifold. We also note that the choice of spin structure is also determined in this frame by the duality chain. Namely, the fermions are periodic in the $\sigma$ direction. 

\subsection{Wavefunctions on the Milne orbifold}
\label{wavefunctions}

Although the frame in which the fermions are periodic on the Milne circle is natural from the perspective of the metric~\C{Milne}, it is not natural for the orbifold description with metric~\C{Mink}. To understand the different approaches, let us turn to a discussion of wavefunctions on Milne space from the perspective of the orbifold construction and associated equivariant bundles. 

The action of the boost \eq{boost} on a wavefunction depends on
the spin $s$. Boost invariant wavefunctions satisfy~\cite{Berkooz:2004re} 
\be\label{periodicity} \phi_s(e^{2\pi
Q\ell_s}\xi^+, e^{-2\pi Q\ell_s}\xi^-)=e^{2\pi
Q\ell_ss}\phi_s(\xi^+,\xi^-). \ee Solutions of the wave equation
satisfying this condition are given by \cite{Berkooz:2004re}
\be\label{solutionphis}
\phi_{j,s}(\xi^+,\xi^-)=\int_{-\infty}^\infty
dv\exp\left[i(p^+ \xi^-e^{-v} +p^-\xi^+e^v)+ivj-vs\right] \ee with
$ Q \ell_{s} j\in\Z$. From the expression \eq{solutionphis}, it follows
that the momentum in the $\sigma$ direction is $j+is$. In
particular, the solutions are not periodic for any non-zero spin.

To gain some intuition for this phenomenon, let us consider the
case of a gauge field, which has spin components $s=\pm1$. The
gauge field can be written as \be A=A_+d\xi^++A_-d\xi^-. \ee
According to \eq{boost}, $A$ will only be boost invariant if \be
A_\pm(e^{2\pi Q\ell_s}\xi^+, e^{-2\pi Q\ell_s}\xi^-)= e^{\mp2\pi
Q\ell_s}A_\pm(\xi^+,\xi^-). \ee Therefore, the components $A_\pm$
satisfy \eq{periodicity} with $s=\mp1$. The reason that the
components $A_\pm$ do not have integer $\sigma$-momenta is that
they are components in a basis $(d\xi^+,d\xi^-)$ that is not
invariant under \eq{boost}.

{}From this discussion, it is clear what we should do if we want
to work with periodic gauge potentials: expand $A$ in a basis of
invariant differential forms. For instance, we can write \be
A=\tilde A_+ dx^++\tilde A_-dx^-. \ee Indeed, one immediately
checks that \be\label{atildea} \tilde
A_\pm={\partial\xi^\pm\over\partial x^\pm}A_\pm=\sqrt2Q\xi^\pm
A_\pm \ee satisfies \be \tilde A_\pm(e^{2\pi Q\ell_s}\xi^+,
e^{-2\pi Q\ell_s}\xi^-)=\tilde A_\pm(\xi^+,\xi^-). \ee

The components $A_\pm$ are those of a gauge field in Minkowski
space \eq{Mink}. In the gauge $\partial^\mu A_\mu=0$, they satisfy
the wave equation \be {\partial^2\over
\partial\xi^+\partial\xi^-}A_\pm=0, \ee which is indeed solved by
\eq{solutionphis} with $p^+p^-=0$. From \eq{atildea}, we find that
$\tilde A_\pm$ satisfy a different wave equation, namely \be
{\partial^2\over \partial x^+\partial x^-}\tilde
A_\pm-\sqrt2Q{\partial\over\partial x^\mp}\tilde A_\pm=0. \ee

In order to discuss fermions, we introduce a vielbein
$e^a_\mu$, such that $\eta_{ab}e^a_\mu e^b_\nu=g_{\mu\nu}$, where
$\mu$ label coordinates $X^\mu$. A choice of vielbein corresponds
to a choice of basis of one-forms \be\label{basisoneforms}
e^a=e^a_\mu dX^\mu. \ee

The kinetic term of a Dirac fermion is given by \be S=\int d^2X
\sqrt g\, [i\bar\psi^\alpha\gamma^\mu_{\a\b}{\cal D}_\mu\psi^\b],
\ee where \be \gamma^\mu_{\a\b}=e^\m_\a\gamma^a_{\a\b}. \ee In
two-dimensional Minkowski space \eq{Mink}, a natural choice of
vielbein is \be\label{vielbeinflat} e^\pm=d\xi^\pm. \ee The
identification \eq{boost} acts on this basis as a non-trivial
Lorentz transformation: \be e^\pm\mapsto e^{\pm 2\pi Q\ell_s}
e^\pm. \ee This Lorentz transformation also acts on the gamma
matrices and thus on the spinor indices $\alpha,\beta$ of the
fermions: \be\label{periodfermion} \psi\mapsto \exp(\pi
Q\ell_s\gamma^{01})\psi, \ee where we can choose a representation
where $\gamma^{01}={\rm diag}(1,-1)$. This implies that the
fermions are not periodic in the $\sigma$ coordinate. The frame
defined by \eq{Mink}\ and \eq{vielbeinflat}\ has the advantage that the
determinant of the metric is constant and that the spin connection
(hidden in ${\cal D}_\m$) vanishes, so that the fermions satisfy a
standard wave equation \be {\partial^2\over
\partial\xi^+\partial\xi^-}\psi=0, \ee which is solved by
\eq{solutionphis} with $p^+p^-=0$.

Alternatively, we can work with the $x^\pm$ coordinates, in terms
of which the metric is given by \eq{Milnebis}, and choose the
vielbein given by \be\label{newvielbein} \tilde
e^\pm=e^{Q(x^++x^-)/\sqrt2}dx^\pm, \ee which is invariant under
the identification. If we define spinor indices using the Lorentz
frame determined by this vielbein, the corresponding spinors will
be periodic in $\sigma$.

Indeed, in terms of the coordinates $x^\pm$, the vielbein
\eq{vielbeinflat} reads \be e^\pm=e^{\sqrt2Qx^\pm}dx^\pm, \ee from
which one obtains the new vielbein \eq{newvielbein} by a Lorentz
transformation that multiplies $e^\pm$ by $\exp(\mp Q\sigma)$.
Therefore, the new fermions $\tilde\psi$ are obtained from the old
ones $\psi$ by the same Lorentz transformation:
\be\label{Lorentzfermion}
\tilde\psi=\exp\left(-{Q\sigma\gamma^{01}\over2}\right)\psi. \ee
>From \eq{periodfermion}, we see indeed that the new fermions
$\tilde\psi$ are periodic in $\sigma$.

Using the vielbein \eq{vielbeinflat}, the spin connection is zero.
The new vielbein \eq{newvielbein} does have a non-trivial spin
connection, \be \omega_{+-}={Q\over\sqrt2}(dx^+-dx^-)=Qd\sigma,
\ee which will appear in the Dirac equation for $\tilde\psi$.  This
Dirac equation can also be derived from the standard Dirac
equation for $\psi$ by substituting \eq{Lorentzfermion}.

\section{The Leading Quantum Mechanical Effects}
\subsection{Some comments on effective actions}
\label{effective}
Matrix string theory~\cite{Dijkgraaf:1997vv, Motl:1997th, Banks:1996my}\ compactified on Milne space is described by the action~\cite{Craps:2005wd}\
\bea
 \label{sym} S &=& {1\over 2\pi \ell_s^2}
  \int {\tr}\Big(  {1\over 2}(D_\mu X^i)^2 + \bar\psi
\Dslash \,\psi + { g_s^2 \ell_s^4 \pi^2} F_{\mu\nu}^2 - {1\over
4 \pi^2 g_s^2\ell_s^4}[X^i,X^j]^2 \cr  && \qquad + {1\over 2\pi g_s\ell_s^2} \bar\psi \gamma_i
[X^i,\psi]\Big)
\eea  
where $g_{s}= e^{-Q\tau}$ and the coordinate $\sigma$ takes values
$
0 \leq \sigma \leq 2\pi \ell_{s}.$ We are using the simpler relation $g_{s}= e^{-Q\tau}$ rather than the exact relation with the space-time parameters, $g_{s}= e^{-{\sqrt{2} \pi \ell_{s} Q\tau \over R} }$, found in~\cite{Craps:2005wd}. This amounts to a redefinition of $Q$ which helps reduce notational clutter. However, if one is interested in the large $N$ limit or the $R$ dependence of the potential then the following replacement
\be\label{replace}
Q \, \rr \,  {\sqrt{2} \pi \ell_{s} Q \over R} 
\ee
should be made in all subsequent formulae. 

 Because of the explicit time-dependence, world-sheet energy is not conserved. In particular, an excitation with fixed energy $E$ is weakly coupled at early times ($\tau \rightarrow -\infty$) since the dimensionful Yang-Mills coupling
\be g_{{YM}}  = {1\over g_{s} \ell_{s}}
\ee
is becoming small. At late times, the theory is strongly coupled and we expect the Yang-Mills theory to flow to type IIA light-cone string field theory in the light-like linear dilaton background. 

In this very late time regime, we have no perturbative control over the theory. However at early times, we can employ perturbation theory to describe the leading quantum mechanical effects. First we note that a vacuum solution is still described by choosing a constant matrix configuration 
in the Cartan of the gauge group. To determine whether there is a potential at $1$-loop, we consider quadratic fluctuations around this vacuum solution. 

Before turning to this computation, it is worth pointing out some subtleties with effective actions in time-dependent backgrounds. The usual procedure involves integrating out massive degrees of freedom. This results in either a 1PI or Wilsonian effective action for the residual light degrees of freedom, depending on the integration technique. However, in a time-dependent background like the Milne orbifold, the Wilsonian procedure always breaks down at sufficiently early times.  

To see this, note from~\C{sym}\ that the mass of the W-bosons is time-dependent: in terms of the SO(8)-invariant distance $b$ between two eigenvalues of $X^i$,
\be
m_{W}^{2} \sim e^{2Q\tau} b^{2}.
\ee
The mass thus vanishes as $\tau \rr -\infty$. Therefore the Wilsonian procedure of integrating out modes with energies above a fixed cut-off $\Lambda$ results in an abelian effective action only for times such that $ m_{W} > \Lambda$. The characteristic break down time is given by
\be
\tau_{\rm non-abelian}  \sim {1\over Q} \ln (\Lambda/b). 
\ee
{}For times earlier than $\tau_{\rm non-abelian}$, the complete non-abelian action should be employed.   

There is a second characteristic time in this system. Namely, the time at which Yang-Mills perturbation theory breaks down. This occurs roughly when $g_{YM}/b \sim 1$ where loops of W-bosons become strongly coupled. This gives a characteristic time 
\be\label{taustring}
\tau_{string} \sim {1\over Q} \ln (\ell_{s} b).
\ee 
At this time, we transition from a Yang-Mills description to a perturbative string description. The perturbative string theory should correspond to the light-cone quantized type IIA string in the light-like linear dilaton background. This string theory becomes more weakly coupled as $\tau \rr \infty$.        

\subsection{The effective action in a loop expansion}

Because of the difficulty in defining a Wilsonian effective action, we will consider the 1PI effective action in a loop expansion. To clarify the expansion parameters, let us rescale the fields in~\C{sym}\ as follows:   
\be\label{redefine}
X^{i}  \rr  \ell_{s}^{2} X^{i} , \qquad
\psi \rr \ell_{s}^{2} \psi,  \qquad
A_{\mu}  \rr  A_{\mu}.
\ee
After this rescaling, the scalars have mass dimension $[X^{i}]=1$ along with the gauge-fields, while the fermions have mass dimension $ [ \psi] = 3/2$.

The rescaled action is given by
\bea
 \label{rescaled} S &=& {\ell_{s}^{2}\over 2\pi }
  \int {\tr}\Big(  {1\over 2}(D_\mu X^i)^2 + \bar\psi
\Dslash \,\psi + { e^{-2Q\tau}  \pi^2} F_{\mu\nu}^2 - {1\over
4 \pi^2 } e^{2Q\tau} [X^i,X^j]^2 \cr  && \qquad+ {1\over 2\pi} e^{Q\tau} \bar\psi \gamma_i
[X^i,\psi]\Big). 
\eea  
There are only two dimensionful parameters in this theory.  We identify $\hbar$ with $1/\ell_{s}^{2}$. This parameter controls the strength of quantum corrections and defines our loop expansion. The second parameter is $Q$. The value of $Q$ has no invariant physical meaning~\cite{Craps:2005wd}: either $Q$ is zero or it is non-zero. If $Q$ is zero, we are back to conventional Matrix string theory on a cylinder with a conventional effective action. 

Around the vacuum discussed in the previous subsection, we have a collection of massless particles corresponding to excitations along the Cartan directions and a collection of particles with time-dependent masses. We would like to integrate out these ``massive'' particles to obtain effective dynamics for the massless degrees of freedom. This is potentially problematic because the time-dependent masses vanish as $\tau \rr -\infty$. Usually, integrating out a massless particle results in a non-local 1PI effective action. This basically comes about because the wave equation for a massless particle is gapless so the corresponding propagator has no analytic expansion in powers of the momentum. To construct the effective action, we usually make use of the expansion
\be
{1\over m^{2}+\Delta} = {1\over m^{2}} \left( 1 - {\Delta\over m^{2}} + \ldots \right)
\ee
to obtain a local action. 
We need to first check whether our particles with time-dependent masses will give rise to the same problems as conventional massless particles. 

The off-diagonal scalar fields have action \be \label{mscalar}
S= {\ell_{s} ^{2}\over 2\pi } \int d\tau
d\sigma\left(\dot{X}^2-{X'}^2- b^{2} e^{2Q \tau} X^2\right). \ee 
The equation of motion is
that of a free scalar with time-dependent mass. The $\sigma$
momentum is quantized in units of $1/\ell_s$. For modes with non-zero $\sigma$ momentum, we
have a conventional mass term and no problem in defining the effective action. So let us consider the case with zero $\sigma$ momentum. We want to study the Green's function for a particle satisfying
\be\label{green}
 \left( \, \p_{\tau}^{2} + b^{2} e^{2Q \tau}  \right) G(\tau,\tau') = \delta(\tau - \tau'). 
\ee  
Note that we are not analytically continuing to Euclidean world-sheet time. The exact propagator for~\C{green}\ can be constructed by patching together linear combinations of the two homogeneous solutions to the equation. These two solutions are given by the Bessel functions $J_{0}(y)$ and $Y_{0}(y)$ where 
\be
y = {b \over Q} e^{Q\tau}. 
\ee
{}For large $y$ or late times, these solutions have good asymptotic behavior
\be
J_{0}(y) \sim \sqrt{{2\over \pi y}} \cos \left(y - {\pi\over 4} \right), \qquad Y_{0}(y) \sim \sqrt{{2\over \pi y}} \sin \left(y - {\pi\over 4} \right).
\ee
This is the behavior we expect for a very massive particle with mass increasing exponentially with time. 

The regime which might be problematic for constructing the 1PI effective action is when $b$ is small, or equivalently, at early times. In this regime, it is easy to construct the propagator in a perturbative expansion around small $b$, 
\bea
\left( \, \p_{\tau}^{2} + b^{2} e^{2Q \tau}  \right)^{-1} & \sim & \int d\omega \, { e^{i \omega (\tau-\tau')} \over -\omega^{2} + b^{2} e^{2Q\tau}} \left\{ 1+ O(b) \right\}, \non \\
& \sim & \left( \alpha_{1} e^{i (\tau - \tau') b e^{Q\tau}} + \alpha_{2} e^{-i (\tau-\tau') b e^{Q\tau}}  \right) \left\{ 1+ O(b) \right\}. 
\eea
The choice of $(\alpha_{1}, \alpha_{2})$ depend on the specific $i\epsilon$ prescription, or equivalently, the specific choice of boundary conditions. The key point is that there appears to be no apparent problem with this propagator. At early times, we can still construct an analytic expansion in $\omega$.  Therefore, we should still be able to construct a 1PI effective action by integrating out particles satisfying~\C{green}.  

The final method of studying quantum corrections makes this conclusion more transparent. Let us switch to $(\xi^{+}, \xi^{-})$ coordinates defined in~\C{relation}. These coordinates are natural for the orbifold description as discussed in section~\ref{wavefunctions}. In this frame, there is no explicit time-dependence in the action. Instead of studying~\C{green}, we end up considering the standard wave-equation for a free massive particle but with an added invariance condition under the orbifold action. In this approach, the difference with flat space Matrix theory computations becomes clearer. All the non-trivial quantum effects come about because of spin-dependent modifications of the $\sigma$ momentum quantization condition. 

{}For example for a scalar particle in this frame, we consider the wave-equation
\be\label{orbwave}
\left( 2  {\partial^2\over
\partial\xi^+\partial\xi^-} + b^{2} \right) X=0 \ee
with the invariance condition given in~\C{periodicity}. In this frame, there is clearly no problem integrating out $X$. 

\subsection{The effective potential}

In this section, we compute the effective potential for the light field $b$, corresponding to the difference between two eigenvalues.
Path-integrating over a massive bosonic field $X$ satisfying~\C{orbwave}\ with constant $b$ gives a determinant
\be\label{determ}
 {\det}^{(-1/2)} \left( H \right)= {\det}^{(-1/2)} \left(  2  {\partial^2\over \partial\xi^+\partial\xi^-} + b^{2}\right). 
\ee
In Feynman diagram language, the determinant \eq{determ} is the sum of all vacuum diagrams, connected or disconnected. The potential term in the effective action is instead the sum of all one-particle-irreducible diagrams in the constant light-field background. At one-loop order, this corresponds to the logarithm
\be
-i\int V_{eff}=\log \det{}^{-1/2}(H)=-{1\over2}\tr\log(H).
\ee
Let us denote the propagator for~\C{orbwave}\ by $G(\xi, \xi'; b^{2})$. We can construct the heat kernel for this wave operator as follows:
\be\label{kernel}
 e^{t H}(\xi, \xi') =   -\oint {dz \over 2 \pi i} \, e^{tz} \, G(\xi, \xi'; b^{2}-z),  
\ee
where the contour integral over $z$ encloses the spectrum of $H$. 
We can then express the determinant in the form
\be \label{potential}
 {\det}^{(-1/2)} \left( H \right) = \exp\left( {1\over 2} \int d^{2} \xi \, \int dt \, {e^{-it(H-i\epsilon)} \over t} (\xi, \xi)\right)
\ee
where an $i\epsilon$ is inserted to ensure convergence at the large $t$ end of the integral. 

In the usual flat space maximally supersymmetric theory, a potential is forbidden by a non-renormalization theorem~\cite{Paban:1998ea}.  At the $1$-loop level, this can be seen explicitly from a cancellation between boson and fermion contributions (see, for example,~\cite{Becker:1997wh}, for the quantum mechanics case).  The count goes as follows: take gauge group $SU(N)$ broken to $U(1)^{(N-1)}$. There are $8 (N^{2}-N)$ massive scalars,  $8 (N^{2}-N)$ massive complex fermions. There are also $(N^{2}-N)$ massive W-bosons and a corresponding $(N^{2}-N)$ complex scalar ghost fields. In standard Matrix theory on a cylinder, the contributions to the potential from these particles cancel. 

To see what happens in this case, let us compute the contribution from a massive spin-$s$ particle.  The fermion satisfies the usual massive wave equation~\C{orbwave}\ except momenta in the $\sigma$ direction are given by
\be
 j\pm s i, \quad Q \ell_{s} j\in\Z
\ee
{}for spin-$s$ particles. 

To complete the computation, we require the Green's functions for these different particles. This requires a choice of vacuum state. There are two natural choices labeled the adiabatic vacuum and the conformal vacuum~\cite{Birrell:1982ix}. The adiabatic vacuum descends from the ambient $1+1$-dimensional Minkowski space under the orbifold identification. We will construct our Green's functions using this choice of vacuum. 

The Green's function for a spin-$s$ particle is obtained by summing over images under the orbifold identification. It takes the form~\cite{Berkooz:2004re}
\be \label{Gs}
G_{s}(\xi, \xi'; b^{2}) =  \sum_{n} \int {dp^{+}dp^{-} \over (2\pi)^{2}} \, {\exp\left({-i p^{-} (\xi^{+} - e^{2\pi Q \ell_{s} n} \xi^{+'}) - i p^{+} (\xi^{-} - e^{-2\pi Q \ell_{s} n} \xi^{-'}) + 2\pi Q \ell_{s} n s} \right) \over - 2 p^{+} p^{-} + b^{2}}. 
\ee
 This gives a kernel
\bea \label{finalker}
  e^{-it H_{s}} (\xi, \xi) & =&   \sum_{n} \int {dp^{+}dp^{-} \over (2\pi)^{2}} \,  \exp\Big(-i t\left[ b^{2} - 2 p^{+} p^{-} \right] \\ \non && + \left[-{i p^{-} \xi^{+}( 1- e^{2\pi Q \ell_{s} n} ) - i p^{+} \xi^{-}(1 - e^{-2\pi Q \ell_{s} n} ) + 2\pi Q \ell_{s} n s} \right]  \Big) \\
&= & \non  \sum_{n} {1\over (2\pi) 2t} \exp\Big(- i t b^{2} - i {\xi^{-} \xi^{+} \over 2 t} ( 1- e^{2\pi Q \ell_{s} n} ) (1 - e^{-2\pi Q \ell_{s} n}) + 2\pi Q \ell_{s} n s  \Big)\\
&= & \non  \sum_{n} {1\over (2\pi) 2t} \exp\Big(- i t b^{2} +2 i {\xi^{-} \xi^{+} \over  t} \sinh^2(\pi Q\ell_s n) + 2\pi Q \ell_{s} n s  \Big)
\eea
with some noteworthy features. If we restrict to $n=0$, this kernel collapses to the usual one for a massive free-particle propagating along a closed loop over time $t$.  This would give rise to the usual Coleman-Weinberg potential. The modification from the free-particle form comes strictly from the orbifold projection in~\C{Gs}\ which introduces space-time dependence into the kernel. 

This space-time dependence has a nice interpretation. The potential~\C{potential}\ based on the kernel~\C{finalker}\ is essentially the Feynman propagator for a particle in $4$ dimensions. In the absence of the $\xi^{-} \xi^{+}$ term,  the usual UV divergences come about from the small $t$ contribution to the integral. The presence of the  $\xi^{-} \xi^{+}$ term regularizes the small $t$ contribution. The interpretation is now in terms of a particle trying to propagate in a small time $t$ a distance squared $\xi^{-} \xi^{+}$ which is exponentially large in $\tau$. Such an amplitude is enormously suppressed as $\tau$ becomes large. On the other hand, the large $t$ contribution to the integral is also suppressed by the mass term proportional to $b^{2}$.        

The kernel \eq{finalker} leads to the following potential term in the action:
\be\label{potentialterm}
\int V_{eff}(b)=i\int d^2\xi\int{dt\over 2t}\sum_{\rm helicities}e^{-it(H_s-i\epsilon)}(\xi,\xi).
\ee
Since ghosts cancel the contributions of two scalars, each supersymmetry multiplet effectively contributes one $s=1$, four $s=1/2$, six $s=0$, four $s=-1/2$ and one $s=-1$ states. Therefore,
\be
\sum_{\rm helicities} (-1)^{2s} \, e^{2\pi Q\ell_ss}=\left(e^{\pi Q\ell_sn/2}-e^{-\pi Q\ell_sn/2}\right)^4=16\sinh^4(\pi Q\ell_sn/2).
\ee
The potential term \eq{potentialterm} thus reads
\be
\int d^2\xi \sum_{n=-\infty}^\infty\left({2i\over\pi}\right)\sinh^4(\pi Q\ell_s n/2)
\int_0^\infty {dt\over t^2}\exp\left(-itb^2+{i\over t}2\sinh^2(\pi Q\ell_s n)\xi^+\xi^-\right).
\ee
Analytically continuing the Schwinger parameter, $t=-it'$, this becomes
\be
- \int d^2\xi \sum_{n=-\infty}^\infty{2\over\pi}\sinh^4(\pi Q\ell_s n/2)
\int_0^\infty {dt^\prime\over (t^\prime)^2}\exp\left(-t^\prime b^2-{1\over t^\prime}2\sinh^2(\pi Q\ell_s n)\xi^+\xi^-\right)
\ee
\be\label{potentialBessel}
= - \int d^2\xi \sum_{n=-\infty}^\infty{2\over\pi}{b\sinh^4(\pi Q\ell_s n/2)\over
[2\sinh^2(\pi Q\ell_sn)\, \xi^+\xi^-]^{1/2}} \;
K_1\!\left(\sqrt{8b^2\sinh^2(\pi Q\ell_sn)\xi^+\xi^-}\right),
\ee
where $K_1$ is a modified Bessel function, with asymptotic behavior
\bea
K_1(z)\approx{1\over \sqrt z}e^{-z}&&\ \ \ \ (z\gg 1);\label{asymplate}\\
K_1(z)\approx {1\over z}&&\ \ \ \ (z\ll 1).\label{asympearly}
\eea

\paragraph{The late time potential:}

{}For very large times of order $\tau_{string}$ defined in~\C{taustring}, our perturbative approximation breaks down. However, we can still ask how our $1$-loop potential behaves in this late time regime. 
To obtain the very late time behavior, we  use the asymptotic behavior \eq{asymplate} to write \eq{potentialBessel} as
\be\label{asympot}
 \int V_{eff}\approx -\int d^2\xi\, {2^{3/4}b^{1/2}\sinh^4(\pi Q\ell_s/2)\over\pi(\xi^+\xi^-)^{3/4}\sinh^{3/2}|\pi Q\ell_s|}\,\exp\left(-\sqrt{8b^2\sinh^2(\pi Q\ell_s)\xi^+\xi^-}\right),
\ee
the dominant contribution coming from $n=\pm1$.
This is hugely suppressed at late times. There is an intuitive way to understand this phenomenon. The extent to which supersymmetry is broken is controlled by the size of the $\sigma$ circle. At $\tau \rr \infty$, the circle becomes large and supersymmetry is effectively restored. 

While there was no a priori reason for us to see the potential vanish at late times at just $1$-loop, it is a result in perfect agreement with the claim that at late times, this theory flows to string field theory in the light-like linear dilaton background~\cite{Craps:2005wd}. In such a theory, there is no perturbative potential. 
However, if we express~\C{asympot}\ in terms of the perturbative string coupling $g_s = e^{-Q \tau}$, 
\be\label{pertpot}
  \int \sqrt{g} \, V_{eff}(b)  \sim \int d\sigma d\tau \, \sqrt{ {b\over g_{s}} }  \exp\left(-{Cb\over g_s} \right)
\ee 
with constant $C$,
we see that there is a non-perturbative potential that appears to be generated by D-branes. 
If higher loop corrections to the potential are more suppressed then we might hope to compute this potential directly in string theory. 

\paragraph{The early time potential:}

{}We have seen that the summand in \eq{potentialBessel} decreases quickly as a function of $n$ when the argument of the modified Bessel function is larger than one. However, at early times, $b^2\xi^+\xi^-\ll1$, the argument is smaller than one for a range of values of $n$, so we should use the asymptotic behavior \eq{asympearly}. We find
\bea
\int V_{eff}&\approx& -\int d^2\xi \sum_{n}{2\over\pi}{b\sinh^4(\pi Q\ell_s n/2)\over
[2\sinh^2(\pi Q\ell_sn)\xi^+\xi^-]^{1/2}} \;
{1\over\sqrt{8b^2\sinh^2(\pi Q\ell_sn)\xi^+\xi^-}}\\ \non
&=& -\int d^2\xi\, {1\over8\pi\xi^+\xi^-}\sum_n\tanh^2(\pi Q\ell_sn/2)\approx \int d^2\xi\, {1\over8\pi^2 Q\ell_s\xi^+\xi^-}\log(2b^2\xi^+\xi^-),
\eea
where the sum was taken over those values of $n$ for which the argument of the modified Bessel function is smaller than 1. At early times, we thus find an attractive one-loop potential between two eigenvalues.

{}Finally, it would be interesting to extend this computation in at least three directions. The first is to extend the computation to higher loops along the lines of~\cite{Becker:1997wh}. This is important in order to understand whether higher loop effects are more suppressed at late times despite the strong coupling. The form of the potential~\C{pertpot}\ suggests that this might be the case with higher loops corresponding to multi-D-brane contributions. 

The second is to go beyond the static potential to moving configurations \`a la~\cite{Douglas:1996yp}. If this static potential is any guide, the structure of the velocity expansion even at order $v^{2}$ should be quite fascinating, and will perhaps shed more light on the gluon phase that replaces the cosmological singularity. 

The third natural direction is to consider the light-like linear dilaton in type IIB string theory using type IIB Matrix theory~\cite{Sethi:1997sw, Banks:1996my}. The potential in this case is likely to have a very different interpretation because of S-duality in space-time. The computation will also be quite different because of possible gauge theory instanton corrections. We might hope that techniques like those used in~\cite{Paban:1998mp}\ could be generalized to obtain an exact answer. 

\section*{Acknowledgements}
It is our pleasure to thank Emil Martinec and Daniel Robbins for helpful conversations. 
The work of B.~C.\ is supported in part by Stichting FOM, by the Belgian Federal Science Policy Office
through the Interuniversity Attraction Pole P5/27, by the European Commission FP6 RTN programme 
MRTN-CT-2004-005104 and by the ``FWO-Vlaanderen'' through project G.0428.06. The work of A.~R.\ 
is supported in part by NSF Grant No.\ PHY-0354993. The work of S.~S.\ is supported in part by NSF 
CAREER Grant No.\ PHY-0094328 and by NSF Grant No.\ PHY-0401814.


\providecommand{\href}[2]{#2}\begingroup\raggedright\endgroup

\end{document}